\documentclass[12pt]{article}
\usepackage[dvips]{graphicx}

\renewcommand{\bar}[1]{\overline{#1}}

\def\ru1{\rule[-0.4truecm]{0mm}{1truecm}}

\begin{document}
\newpage

%\begin{flushright}
%October 2013
%\end{flushright}

\bigskip

%{\centerline{\Large
%\bf Origin of the Discrepancy between Quark}}
{\centerline{\Large
\bf Quark Helicity and Transversity Distributions}}

\vspace{1.5cm}

\centerline{
\bf Dae Sung Hwang}

\vspace{4mm} \centerline{\it Institute of Fundamental Physics, Sejong
University, Seoul 05006, South Korea}

\centerline{e-mail: dshwang@sejong.edu}

\vspace*{2.5cm}

\begin{abstract}
\noindent
The quark transversity distribution inside nucleon is less understood
than the quark unpolarized and helicity distributions inside nucleon.
In particular, it is important to know clearly why the quark helicity
and transversity distributions are different.
We investigate the origin of their discrepancy.
\end{abstract}

\vfill

\noindent
PACS numbers: 12.39.-x, 12.39.Ki, 14.20.-c, 14.20.Dh\\
Key words: Helicity Distribution, Transversity Distribution, Quark Structure inside Nucleon

\newpage

\section{Introduction}

The unpolarized distribution $f_1(x)$ and the helicity distribution $g_1(x)$
of quarks inside nucleon have been extensively investigated.
However, the transversity distribution $h_1(x)$ is less known since it can not be measured in
fully deep inelastic scattering since it is chiral-odd.
The transversity distribution $h_1(x)$ can be extracted by measuring the double-spin asymmetry
in the Drell-Yan process $A^{\uparrow}\, B^{\uparrow}\to l^+\, l^-\, X$,
where $A^{\uparrow}$ and $B^{\uparrow}$ are two transversely polarized protons or antiprotons,
$l^+\, l^-$ are lepton pairs and $X$ is the undetected hadronic system \cite{BDR02}.
There is also an approach which applies the Collins mechanism \cite{Collins:1992kk} to
the single-spin asymmetry in the process $lp^{\uparrow}\to l\pi X$ of
semi-inclusive deep inelastic scattering.
By these methods important experimental and theoretical progresses have been made
in investigating the quark transversity distribution inside nucleon
\cite{Seidl:2008xc,Airapetian:2010ds,Adolph:2014zba,Anselmino:2013vqa}.

The transversity distribution was first introduced in Ref. \cite{Ralston:1979ys}, and then
there have been extensive studies on this subject
\cite{BDR02,Artru:1989zv,Jaffe:1991kp,Jaffe:1991ra,Jaffe:1996zw}.
However, it is desirable to have a better understanding of the origin of the difference between
$g_1(x)$ and $h_1(x)$.
For example, we can find the following sentence in Ref. \cite{Jaffe:1996zw}:
``It would be very useful to have a better idea of the dynamical and relativistic effects
which generate differences between $g_1$ and $h_1$.''
In this paper we show that the discrepancy between the helicity and
transversity distributions is rooted in the difference between the
Bjorken-Drell spinors and the light-front spinors.
As a result, the quark helicity and transversity distributions are equal if the quark
has no transverse momentum, since the Bjorken-Drell spinors and the light-front spinors
are the same when the transverse momentum is zero.

In this paper we show that the precise description of $f_1(x)$, $g_1(x)$ and $h_1(x)$ is as follows:
In a hadron $f_1(x)$ is the probability of finding a quark with momentum
fraction $x$ of the longitudinal momentum of the hadron.
In a longitudinally polarized hadron $g_1(x)$ is the number density of quarks with momentum
fraction $x$ and in the spin state of $u_1^{\rm LF}(k)$ minus the number density of quarks with the same
momentum fraction and in the spin state of $u_2^{\rm LF}(k)$.
In a hadron transversely polarized to the positive $x$ direction
$h_1(x)$ is the number density of quarks with momentum
fraction $x$ and in the spin state of ${1\over {\sqrt{2}}}(u_1^{\rm LF}(k)+u_2^{\rm LF}(k))$
minus the number density of quarks with the same momentum fraction and in the spin state of
${1\over {\sqrt{2}}}(u_1^{\rm LF}(k)-u_2^{\rm LF}(k))$,
where $u_1^{\rm LF}(k)$ and $u_2^{\rm LF}(k)$ are the light-front spinors defined in Eq. (\ref{s2}).

In the literature the transversity distribution $h_1(x)$ is commonly described by an expression like
``In a transversely polarised hadron $h_1(x)$ is the number density of quarks with momentum
fraction $x$ and polarization parallel to that of the hadron minus the number density of quarks
with the same momentum fraction and antiparallel polarization.''
In this description it is not clear what is meant by ``polarization parallel to that of the hadron''.
We described $h_1(x)$ in the later part of the previous paragraph without such ambiguity.
In reality the quark spin state of ${1\over {\sqrt{2}}}(u_1^{\rm LF}(k) + u_2^{\rm LF}(k))$ is not
the spin state polarized along the positive $x$ direction.
In addition it is not accurate to call $g_1(x)$ the helicity distribution, since the helicity eigenstate
spinors and the light-front spinors are the same only when the quark mass is zero.
In this paper we explain these properties and show why $g_1(x)$ and $h_1(x)$ are different
and when they are equal.

%\vfill\pagebreak

\section{Parton Distributions}

\subsection{Definitions}

The transverse momentum dependent parton distributions
are defined through the vector, axial-vector and tensor currents:
\begin{eqnarray}
&&\int\frac{d y^-d^2{\vec y}_{\perp}}{16(\pi)^3}\;
e^{ix P^+y^--i{\vec k}_{\perp}\cdot {\vec y}_{\perp}}\;
\langle P,{\lambda}' | \bar\psi(0)\,\gamma^+\,\psi(y)\,|P,{\lambda}
\rangle
\Big|_{y^+=0}
\nonumber\\
&&\qquad =\
{1\over 2 P^+}\ {\bar U}(P,{\lambda}') \ f_1(x,{\vec k}_{\perp})
\ {\gamma^+} \ U(P,{\lambda})\ ,
\nonumber\\
&&\int\frac{d y^-d^2{\vec y}_{\perp}}{16(\pi)^3}\;
e^{ix P^+y^--i{\vec k}_{\perp}\cdot {\vec y}_{\perp}}\;
\langle P,{\lambda}' | \bar\psi(0)\,\gamma^+\gamma_5\,\psi(y)\,|P,{\lambda}
\rangle
\Big|_{y^+=0}
\nonumber\\
&&\qquad =\
{1\over 2 P^+}\ {\bar U}(P,{\lambda}') \ g_1(x,{\vec k}_{\perp})
\ {\gamma^+}\gamma_5 \ U(P,{\lambda})\ ,
\nonumber\\
&&\int\frac{d y^-d^2{\vec y}_{\perp}}{16(\pi)^3}\;
e^{ix P^+y^--i{\vec k}_{\perp}\cdot {\vec y}_{\perp}}\;
\langle P,{\lambda}' | \bar\psi(0)\,\sigma^{+i}\,\psi(y)\,|P,{\lambda}
\rangle
\Big|_{y^+=0}
\nonumber\\
&&\qquad =\
{1\over 2 P^+}\ {\bar U}(P,{\lambda}') \ h_1(x,{\vec k}_{\perp})
\ \sigma^{+i} \ U(P,{\lambda})\ .
\label{pa1}
\end{eqnarray}
The parton distributions
$f_1(x)$, $g_1(x)$, $h_1(x)$ are given by integrating the unintegrated
parton distributions over ${\vec k}_{\perp}$:
\begin{eqnarray}
f_1(x)&=&
\int \Big[ {\mathrm d}^2{\vec k}_{\perp}\Big]\
f_1(x,{\vec k}_{\perp})\ ,
\nonumber\\
g_1(x)&=&
\int \Big[ {\mathrm d}^2{\vec k}_{\perp}\Big]\
g_1(x,{\vec k}_{\perp})\ ,
\nonumber\\
h_1(x)&=&
\int \Big[ {\mathrm d}^2{\vec k}_{\perp}\Big]\
h_1(x,{\vec k}_{\perp})
\ ,
\label{pa4intdk2}
\end{eqnarray}
where $\Big[ {\mathrm d}^2{\vec k}_{\perp}\Big]$ is ${\mathrm d}^2{\vec k}_{\perp}$
times a common overall constant which normalizes $f_1(x)$ to satisfy
$\int_0^1\, f_1(x)\, dx=1$.

\subsection{Wavefunction Representations}

The state of proton
is represented by the light-front Fock expansion \cite{BHMS, BDH}:
\begin{eqnarray}
\left\vert \psi_p(P^+, {\vec P_\perp}; \lambda )\right>
&=& \sum_{n}\
\prod_{i=1}^{n}
{{\rm d}x_i\, {\rm d}^2 {\vec k_{\perp i}}
\over \sqrt{x_i}\, 16\pi^3}\ \,
16\pi^3 \delta\left(1-\sum_{i=1}^{n} x_i\right)\,
\delta^{(2)}\left(\sum_{i=1}^{n} {\vec k_{\perp i}}\right)
\nonumber
\\
&& \qquad \rule{0pt}{4.5ex}
\times \psi_n^{\lambda}(x_i,{\vec k_{\perp i}},
\lambda_i) \left\vert n;\,
x_i P^+, x_i {\vec P_\perp} + {\vec k_{\perp i}}, \lambda_i\right>\ ,
\label{pa3}
\end{eqnarray}
where $x_i = k^+_i/P^+$ and ${\vec k_{\perp i}}$ is the relative
transverse momentum of constituent.
From (\ref{pa1}) and (\ref{pa3}) we find that
the transverse momentum dependent parton distributions
are expressed in terms of the light-front wavefunctions as
\begin{eqnarray}
f_1(x,{\vec k}_{\perp})
&=&
{\cal A}\
\psi^{\uparrow \ *}_{(n)}(x_i,
  {\vec{k}}_{\perp i},\lambda_i) \
\psi^{\uparrow}_{(n)}(x_i, {\vec{k}}_{\perp i},\lambda_i) \ ,
\nonumber\\
&=&
{\cal A}\
\psi^{\downarrow \ *}_{(n)}(x_i,
{\vec{k}}_{\perp i},\lambda_i) \
\psi^{\downarrow}_{(n)}(x_i, {\vec{k}}_{\perp i},\lambda_i) \ ,
\nonumber\\
g_1(x,{\vec k}_{\perp})
&=&
{\cal A}\
\lambda_1\
\psi^{\uparrow \ *}_{(n)}(x_i,
  {\vec{k}}_{\perp i},\lambda_i) \
\psi^{\uparrow}_{(n)}(x_i, {\vec{k}}_{\perp i},\lambda_i) \ ,
\nonumber\\
&=&
{\cal A}\
(-\lambda_1)\
\psi^{\downarrow \ *}_{(n)}(x_i,
{\vec{k}}_{\perp i},\lambda_i) \
\psi^{\downarrow}_{(n)}(x_i, {\vec{k}}_{\perp i},\lambda_i) \ ,
\nonumber\\
h_1(x,{\vec k}_{\perp})
&=&
{\cal A}\
\psi^{\downarrow \ *}_{(n)}(x_i,
{\vec{k}}_{\perp i},
{\lambda}^\prime_{1}={\downarrow},\lambda_{i{\ne}1}) \
\psi^{\uparrow}_{(n)}(x_i, {\vec{k}}_{\perp i},
{\lambda}_{1}={\uparrow},\lambda_{i{\ne}1})
\ ,
\nonumber\\
&=&
{\cal A}\
\psi^{\uparrow \ *}_{(n)}(x_i,
{\vec{k}}_{\perp i},
{\lambda}^\prime_{1}={\uparrow},\lambda_{i{\ne}1}) \
\psi^{\downarrow}_{(n)}(x_i, {\vec{k}}_{\perp i},
{\lambda}_{1}={\downarrow},\lambda_{i{\ne}1})
\ ,
\label{pa4}
\end{eqnarray}
where
\begin{equation}
{\cal A}\ =\
\sum_{n, \lambda_i}
\int \prod_{i=1}^{n}
{{\rm d}x_{i}\, {\rm d}^2{\vec{k}}_{\perp i} \over 16\pi^3 }\ \,
16\pi^3 \delta\left(1-\sum_{j=1}^n x_j\right) \, \delta^{(2)}
\left(\sum_{j=1}^n {\vec{k}}_{\perp j}\right)\
\delta(x-x_{1})\ \delta^{(2)}({\vec{k}}_{\perp}-{\vec{k}}_{\perp 1})\ .
\label{pa5}
\end{equation}

The formulas given in (\ref{pa4}) can be used to find the transverse
momentum dependent distributions in an explicit model.
These formulas can also be applied in getting model independent
relations. For example, we can show the Soffer's inequality \cite{soffer95}.
After some calculations, from (\ref{pa4}) we get
\begin{eqnarray}
&&\ \Big[ \ \Bigl( f_1(x,{\vec k}_{\perp})+g_1(x,{\vec k}_{\perp})\Bigr)\
\pm\ 2\ h_1(x,{\vec k}_{\perp})\ \Big]
\ =\ {\cal A}
\nonumber\\
%&& \qquad \rule{0pt}{3ex} {} \times\
&& \times\
\Big[\
\psi^{\uparrow \ *}_{(n)}(x_i,
  {\vec{k}}_{\perp i},{\lambda}^\prime_{1}={\uparrow},\lambda_{i{\ne}1}) \
\pm\
\psi^{\downarrow \ *}_{(n)}(x_i, {\vec{k}}_{\perp i}
{\lambda}^\prime_{1}={\downarrow},\lambda_{i{\ne}1})\ \Big]
\nonumber\\
%&& \qquad \rule{0pt}{3ex} {} \times\
&& \times\
\Big[\
\psi^{\uparrow}_{(n)}(x_i,
  {\vec{k}}_{\perp i},{\lambda}_{1}={\uparrow},\lambda_{i{\ne}1}) \
\pm\
\psi^{\downarrow}_{(n)}(x_i, {\vec{k}}_{\perp i}
{\lambda}_{1}={\downarrow},\lambda_{i{\ne}1})\ \Big]
\ ,
\label{pa6}
\end{eqnarray}
which shows the Soffer's inequality as
\begin{equation}
\Bigl( f_1(x,{\vec k}_{\perp})+g_1(x,{\vec k}_{\perp})\Bigr)\
\pm\ 2\ h_1(x,{\vec k}_{\perp})\ \ge\ 0\ ,
\label{pa7}
\end{equation}
where the equality holds when
\begin{equation}
\psi^{\uparrow}_{(n)}(x_i,
  {\vec{k}}_{\perp i},{\lambda}_{1}={\uparrow},\lambda_{i{\ne}1}) \
\pm\
\psi^{\downarrow}_{(n)}(x_i, {\vec{k}}_{\perp i}
{\lambda}_{1}={\downarrow},\lambda_{i{\ne}1})\ =\ 0\ .
\label{pa8}
\end{equation}

From the formulas for $f_1(x)$, $g_1(x)$ and $h_1(x)$ given by Eqs. (\ref{pa4intdk2}) and (\ref{pa4}),
we can show the following:
In a hadron $f_1(x)$ is the probability of finding a quark with momentum
fraction $x$ of the longitudinal momentum of the hadron.
In a longitudinally polarized hadron $g_1(x)$ is the number density of quarks with momentum
fraction $x$ and in the spin state of $u_1^{\rm LF}(k)$ minus the number density of quarks with the same
momentum fraction and in the spin state of $u_2^{\rm LF}(k)$.
In a hadron transversely polarized to the positive $x$ direction
$h_1(x)$ is the number density of quarks with momentum
fraction $x$ and in the spin state of ${1\over {\sqrt{2}}}(u_1^{\rm LF}(k)+u_2^{\rm LF}(k))$
minus the number density of quarks with the same momentum fraction and in the spin state of
${1\over {\sqrt{2}}}(u_1^{\rm LF}(k)-u_2^{\rm LF}(k))$,
where $u_1^{\rm LF}(k)$ and $u_2^{\rm LF}(k)$ are the light-front spinors given in Eq. (\ref{s2}).
We emphasize that
the quark spin states of $u_1^{\rm LF}(k)$ and $u_2^{\rm LF}(k)$ are not
the quark helicity eigenstates when quark mass is not zero as we can see
in Eqs. (\ref{s13bapp}) and (\ref{s15bapp}) in Appendix A,
and the quark spin states of ${1\over {\sqrt{2}}}(u_1^{\rm LF}(k)\pm u_2^{\rm LF}(k))$ are not
the angular momentum eigenstates polarized along the $\pm x$ directions.
Those eigenstates are
the quark spin states of ${1\over {\sqrt{2}}}(u_1^{\rm BD}(k)\pm u_2^{\rm BD}(k))$,
where $u_1^{\rm BD}(k)$ and $u_2^{\rm BD}(k)$ are the Bjorken-Drell spinors given in Eq. (\ref{s1}).
We show this property in Appendix B.

\section{The reason why $g_1(x)\ne h_1(x)$}

We use
the notations
$k^R=k^1+ik^2$,
$k^L=k^1-ik^2$,
$k^{\pm}=k^0\pm k^3$,
and the $\gamma$ matrices in the Dirac representation:
\begin{equation}
\gamma^0=\left(
\begin{array}{cc}
1&0\\
0&-1
\end{array}
\right)
\ ,\ \
\gamma^i=\left(
\begin{array}{cc}
0&{\sigma}^i\\
-{\sigma}^i&0
\end{array}
\right)
\ ,\ \
\gamma_5=\left(
\begin{array}{cc}
0&1\\
1&0
\end{array}
\right)
\ ,
\label{gammamat}
\end{equation}
where ${\sigma}^i$ are the Pauli matrices given in Appendix A.

We consider two sets of the positive energy solutions of the Dirac equation
\begin{equation}
({\not{k}}-m)\ u(k)\ =0\ .
%{\not{k}}u(k)=mu(k)\ .
\label{diraceq}
\end{equation}
The Bjorken-Drell spinors are two linearly independent solutions
of (\ref{diraceq}) \cite{BD64} given by
\begin{equation}
u_1^{\rm BD}(k)={1\over {\sqrt{k^0+m}}}
\left(
\begin{array}{c}
k^0+m\\
0\\
k^3\\
k^R
\end{array}
\right)
\ ,
\qquad
u_2^{\rm BD}(k)={1\over {\sqrt{k^0+m}}}
\left(
\begin{array}{c}
0\\
k^0+m\\
k^L\\
-k^3
\end{array}
\right)
\ .
\label{s1}
\end{equation}
The light-front spinors are another set of linear combinations of
the solutions of (\ref{diraceq}) \cite{KS70,LB80,BPP98}
given, in the convention of Ref. \cite{LB80}, by
\begin{equation}
u_1^{\rm LF}(k)={1\over {\sqrt{2k^+}}}
\left(
\begin{array}{c}
k^++m\\
k^R\\
k^+-m\\
k^R
\end{array}
\right)
\ ,
\qquad
u_2^{\rm LF}(k)={1\over {\sqrt{2k^+}}}
\left(
\begin{array}{c}
-k^L\\
k^++m\\
k^L\\
-k^++m
\end{array}
\right)
\ .
\label{s2}
\end{equation}
The two sets $u^{\rm BD}(k)$ in (\ref{s1}) and
$u^{\rm LF}(k)$ in (\ref{s2}) are related as:
\begin{eqnarray}
u_1^{\rm BD}(k)&=&{1\over {\sqrt{(k^++m)^2+{\vec k}_{\perp}^2}}}
\ \Bigl(\
(k^++m)\ u_1^{\rm LF}(k)\ -\
k^R\ u_2^{\rm LF}(k)\ \Bigr)\ ,
\nonumber\\
u_2^{\rm BD}(k)&=&{1\over {\sqrt{(k^++m)^2+{\vec k}_{\perp}^2}}}
\ \Bigl(\
k^L\ u_1^{\rm LF}(k)\ +\
(k^++m)\ u_2^{\rm LF}(k)\ \Bigr)\ .
\label{s4}
\end{eqnarray}
We organize the relations among the Bjorken-Drell spinors, the light-front
spinors and the helicity eigenstate spinors in Appendix A.
%\subsection{The reason why $g_1(x)\ne h_1(x)$}

The Bjorken-Drell spinors
$u_{1,2}^{\rm BD}(k)$ given in (\ref{s1}) satisfy
\begin{equation}
j^+u_1^{\rm BD}(k)=0\ ,\ \ \
j^-u_2^{\rm BD}(k)=0\ ,\ \ \
j^-u_1^{\rm BD}(k)=u^{\rm BD}_2(k)\ ,\ \ \
j^+u_2^{\rm BD}(k)=u_1^{\rm BD}(k)\ ,
\label{p4}
\end{equation}
where
$j^{\pm}=j^1\pm j^2$ and
\begin{equation}
j^i = s^i + l^i\ ,\ \ \
s^i={1\over 2}\, \Sigma^i =
{1\over 2}
\left(
\begin{array}{cc}
\sigma^i&0\\
0&\sigma^i
\end{array}
\right)
\ ,\ \ \
l^i = -i \epsilon^{ijk} k^j{\partial\over \partial k^k}\ ,\ \ \
\epsilon^{123}=1\ .
\label{p5}
\end{equation}
We explain in Appendix B the reason why Bjorken-Drell spinors satisfy (\ref{p4}).

Since $u_{1,2}^{\rm BD}(k)$ satisfy the transformation properties (\ref{p4}),
they are spin-half states which are eigenstates of $j^3$ with eigenvalues $\pm {1\over 2}$ and
the ${1\over {\sqrt{2}}}(u_1^{\rm BD}(k) \pm u_2^{\rm BD}(k))$ state are spin-half states which are
eigenstates of $j^2$ with the eigenvalues $\pm {1\over 2}$.
Then we can construct the proton spin states $|P; \lambda =\pm {1\over 2}>$
by using the Clebsch-Gordan coefficients with $u_{1,2}^{\rm BD}(k)$ for 
the quark state,
which we will do in the next section.
The proton states $|P; \lambda =\pm {1\over 2}>$ satisfy
\begin{eqnarray}
&&{J}^+|P; \lambda ={1\over 2}>=0\ ,\ \ \
{J}^-|P; \lambda =-{1\over 2}>=0
\nonumber\\
&&{J}^-|P; \lambda ={1\over 2}>=|P; \lambda =-{1\over 2}>\ ,
\ \ \
{J}^+|P; \lambda =-{1\over 2}>=|P; \lambda ={1\over 2}>\ ,
\label{p4a}
\end{eqnarray}
where ${J}^i$ is the total angular momentum operator for the proton given by
${J}^i=\sum_a j^i_a =\sum_a ({s}^i_a+{l}^i_a)$,
which is the sum over the constituents $a$,
and $J^{\pm}=J^1\pm J^2$.

We define $g_1^{\rm BD}(x)$ as the probability of the quark's being
in the $u_1^{\rm BD}(k)$ state minus that of being in the
$u_2^{\rm BD}(k)$ state when the proton's state is $|\lambda =+{1\over 2}>$,
where the $u_1^{\rm BD}(k)$, $u_2^{\rm BD}(k)$ and $|\lambda =+{1\over 2}>$ states
are all angular momentum eigenstates of spin-half with the eigenvalues of $j^3$ (or $J^3$)
as $+{1\over 2}$ or $-{1\over 2}$.
We define $h_1^{\rm BD}(x)$ as the probability of the quark's being
in the ${1\over {\sqrt{2}}}(u_1^{\rm BD}(k)+u_2^{\rm BD}(k))$ state minus
that of being in the ${1\over {\sqrt{2}}}(u_1^{\rm BD}(k)-u_2^{\rm BD}(k))$ state
when the proton's state is ${1\over {\sqrt{2}}}(|\lambda =+{1\over 2}>+|\lambda =-{1\over 2}>)$.
Here, $|\lambda =\pm {1\over 2}>$ means $|P; \lambda =\pm {1\over 2}>$.
When the proton spin states $|P; \lambda >$ satisfy (\ref{p4a}),
${1\over {\sqrt{2}}}(|\lambda =+{1\over 2}>+|\lambda =-{1\over 2}>)$ is
an angular momentum eigenstate of spin-half with the eigenvalue of $J^2$ as $+{1\over 2}$,
whereas the ${1\over {\sqrt{2}}}(u_1^{\rm BD}(k)+u_2^{\rm BD}(k))$ and
${1\over {\sqrt{2}}}(u_1^{\rm BD}(k)-u_2^{\rm BD}(k))$ states are
angular momentum eigenstates of spin-half with the eigenvalues of $j^2$ as
$+{1\over 2}$ and $-{1\over 2}$, respectively.
%These ${1\over {\sqrt{2}}}(|\lambda =+{1\over 2}>+|\lambda =-{1\over 2}>)$,
%${1\over {\sqrt{2}}}(u_1^{\rm BD}(k)+u_2^{\rm BD}(k))$ and
%${1\over {\sqrt{2}}}(u_1^{\rm BD}(k)-u_2^{\rm BD}(k))$
%states can be obtained by rotating
%the $|\lambda =+{1\over 2}>$, $u_1^{\rm BD}(k)$ and $u_2^{\rm BD}(k)$ states 90 degrees
%along the positive $y$ axis.
Therefore, the former and latter three states are equivalent
and only their quantization axes are different.
Then, the relation $g_1^{\rm BD}(x)=h_1^{\rm BD}(x)$ is satisfied.

On the other hand, the helicity and transversity distributions $g_1(x)$
and $h_1(x)$ given by Eqs. (\ref{pa4intdk2}) and (\ref{pa4})
can be interpreted as:
$g_1(x)$ is the probability of the quark's being
in the $u_1^{\rm LF}(k)$ state minus that of being in the
$u_2^{\rm LF}(k)$ state when the proton's state is $|\lambda =+{1\over 2}>$,
and $h_1(x)$ is the probability of the quark's being
in the ${1\over {\sqrt{2}}}(u_1^{\rm LF}(k)+u_2^{\rm LF}(k))$ state minus
that of being in the ${1\over {\sqrt{2}}}(u_1^{\rm LF}(k)-u_2^{\rm LF}(k))$ state
when the proton's state is ${1\over {\sqrt{2}}}(|\lambda =+{1\over 2}>+|\lambda =-{1\over 2}>)$.
%which is an eigenstate of $J^2$ with eigenvalue $+{1\over 2}$
%since the proton's states satisfy (\ref{p4a}).
That is, $g_1(x)$ and $h_1(x)$ are related with the light-front spinors.
On the contrary to the case of the Bjorken-Drell spinors,
the light-front spinors given in (\ref{s2}) do not satisfy the property (\ref{p4})
and then the ${1\over {\sqrt{2}}}(u_1^{\rm LF}(k) \pm u_2^{\rm LF}(k))$ states
are not eigenstates of $j^2$,
whereas the ${1\over {\sqrt{2}}}(|\lambda =+{1\over 2}>+|\lambda =-{1\over 2}>)$ proton state
is an eigenstate of $J^2$ with eigenvalue $+{1\over 2}$.
Therefore, the situation for $g_1(x)$ and that for $h_1(x)$ are not equivalent, and then
$g_1(x)$ and $h_1(x)$ are different.
We will see these properties in explicit examples in the next section.
When the transverse momentum is zero, the Bjorken-Drell spinors and
the light-front spinors are the same as we can see in (\ref{s4}).
Therefore, when the quark transverse momentum is zero, $g_1(x)$ and $h_1(x)$ are equal.

\vfill\pagebreak

\section{Explicit Calculations in Diquark Models}

In this section we perform explicit calculations in diquark models in order
to see by examples what we found in previous sections.
We use the Bjorken-Drell spinors for the quark spin states when we construct the nucleon spin states
by using the Clebsch-Gordan coefficients, and then the resulting nucleon spin states
become eigenstates of the total angular momentum.

\subsection{S-wave Scalar Diquark Model}

The nucleon state composed of a scalar diquark and an S-wave quark is
represented as
\begin{eqnarray}
|\uparrow >_N&=&
|\uparrow >_q^{\rm BD}\ {\sqrt{1\over 4\pi}}\
R_0(|{\vec k}|)
\ ,
\nonumber\\
|\downarrow >_N&=&
|\downarrow >_q^{\rm BD}\ {\sqrt{1\over 4\pi}}\
R_0(|{\vec k}|)
\ .
\label{downN0}
\end{eqnarray}
The nucleon state represented by (\ref{downN0}) have the distribution
functions given by
\begin{eqnarray}
f_1(x)&=&
\int \Big[ {\mathrm d}^2{\vec k}_{\perp}\Big]
{1\over 4\pi}
\Big( R_0(|{\vec k}|){\Big)}^2 \ ,
\nonumber\\
%\label{k10}\\
g_1(x)&=&
\int \Big[ {\mathrm d}^2{\vec k}_{\perp}\Big]
{1\over 4\pi}
\Big( R_0(|{\vec k}|){\Big)}^2
\ {1\over (k^++m)^2+{\vec k}_{\perp}^2}\
\Big[ (k^++m)^2-{\vec k}_{\perp}^2\Big] \ ,
\nonumber\\
%\label{k20}\\
h_1(x)&=&
\int \Big[ {\mathrm d}^2{\vec k}_{\perp}\Big]
{1\over 4\pi}
\Big( R_0(|{\vec k}|){\Big)}^2
\ {1\over (k^++m)^2+{\vec k}_{\perp}^2}\
\Big[ (k^++m)^2\Big] \ ,
\nonumber\\
%\label{k30}\\
g^{\rm BD}_1(x)&=&
\int \Big[ {\mathrm d}^2{\vec k}_{\perp}\Big]
{1\over 4\pi}
\Big( R_0(|{\vec k}|){\Big)}^2
\ .
\label{k40}
\end{eqnarray}
We checked by explicit calculation
that $h^{\rm BD}_1(x,{\vec k}_{\perp})$ is the same as
$g^{\rm BD}_1(x,{\vec k}_{\perp})$ given in (\ref{k40}).
{}From the results in (\ref{k40}), we see that the Soffer's
inequality is satisfied with equality in this model:
$f_1(x)+g_1(x)=2h_1(x)$.

\subsection{S-wave Axial-vector Diquark Model}

The nucleon state composed of an axial-vector diquark and an S-wave quark is
represented as
\begin{eqnarray}
|\uparrow >_N&=&\Big(
-{\sqrt{1\over 3}}\ |\uparrow >_q^{\rm BD}\ |10>_{av}
+{\sqrt{2\over 3}}\ |\downarrow >_q^{\rm BD}\ |1+1>_{av}
\Big)\ {\sqrt{1\over 4\pi}}\ R_0(|{\vec k}|)
\ ,
\nonumber\\
%\label{upNv}\\
|\downarrow >_N&=&\Big(
-{\sqrt{2\over 3}}\ |\uparrow >_q^{\rm BD}\ |1-1>_{av}
+{\sqrt{1\over 3}}\ |\downarrow >_q^{\rm BD}\ |10>_{av}
\Big)\ {\sqrt{1\over 4\pi}}\ R_0(|{\vec k}|)
\ .
\label{downNv}
\end{eqnarray}
The nucleon state represented by (\ref{downNv}) have the distribution
functions given by
\begin{eqnarray}
f_1(x)&=&
\int \Big[ {\mathrm d}^2{\vec k}_{\perp}\Big]
{1\over 4\pi}
\Big( R_0(|{\vec k}|){\Big)}^2 \ ,
\nonumber\\
g_1(x)&=&-{1\over 3}
\int \Big[ {\mathrm d}^2{\vec k}_{\perp}\Big]
{1\over 4\pi}
\Big( R_0(|{\vec k}|){\Big)}^2
\ {1\over (k^++m)^2+{\vec k}_{\perp}^2}\
\Big[ (k^++m)^2-{\vec k}_{\perp}^2\Big] \ ,
\nonumber\\
h_1(x)&=&-{1\over 3}
\int \Big[ {\mathrm d}^2{\vec k}_{\perp}\Big]
{1\over 4\pi}
\Big( R_0(|{\vec k}|){\Big)}^2
\ {1\over (k^++m)^2+{\vec k}_{\perp}^2}\
\Big[ (k^++m)^2\Big] \ ,
\nonumber\\
g^{\rm BD}_1(x)&=&-{1\over 3}
\int \Big[ {\mathrm d}^2{\vec k}_{\perp}\Big]
{1\over 4\pi}
\Big( R_0(|{\vec k}|){\Big)}^2
\ .
\label{k4v}
\end{eqnarray}
We checked by explicit calculation
that $h^{\rm BD}_1(x,{\vec k}_{\perp})$ is the same as
$g^{\rm BD}_1(x,{\vec k}_{\perp})$ given in (\ref{k4v}).
{}From the results in (\ref{k4v}), we see that the Soffer's
inequality $f_1(x)+g_1(x)>2|h_1(x)|$ is satisfied.

\subsection{P-wave Scalar Diquark Model}

In this section we consider the P-wave scalar diquark model,
in which the orbital angular momentum of the quark 
is incorporated in the spin contents of nucleon.
We consider here the scalar diquark to be a pseudo-scalar one
in order that the parity of the nucleon is even.
Following the usual construction,
the nucleon state composed of a scalar diquark and a P-wave quark is
represented as \cite{GA06}
\begin{eqnarray}
|\uparrow >_N&=&\Big(
-{\sqrt{1\over 3}}\ |\uparrow >_q^{\rm BD}Y_{10}({\hat k})
+{\sqrt{2\over 3}}\ |\downarrow >_q^{\rm BD}Y_{1+1}({\hat k})
\Big)\ R_1(|{\vec k}|)
\ ,
\nonumber\\
|\downarrow >_N&=&\Big(
-{\sqrt{2\over 3}}\ |\uparrow >_q^{\rm BD}Y_{1-1}({\hat k})
+{\sqrt{1\over 3}}\ |\downarrow >_q^{\rm BD}Y_{10}({\hat k})
\Big)\ R_1(|{\vec k}|)
\ ,
\label{downN}
\end{eqnarray}
where
\begin{equation}
Y_{10}({\hat k})={\sqrt{3\over 4\pi }}\ {k^3\over |{\vec k}|}\ ,
\qquad
Y_{1\pm 1}({\hat k})=\mp {\sqrt{3\over 8\pi }}\ 
{k^1\pm ik^2\over |{\vec k}|}\ .
\label{Y1m}
\end{equation}
The nucleon state represented by (\ref{downN}) have the distribution
functions given by
\begin{eqnarray}
f_1(x)&=&
\int \Big[ {\mathrm d}^2{\vec k}_{\perp}\Big]
{1\over 4\pi}
\Big( R_1(|{\vec k}|){\Big)}^2
\ {(k^0+m)^2\over (k^++m)^2+{\vec k}_{\perp}^2}\
{1\over |{\vec k}|^2}\Big[ (k^+-m)^2+{\vec k}_{\perp}^2\Big] \ ,
\nonumber\\
g_1(x)&=&
\int \Big[ {\mathrm d}^2{\vec k}_{\perp}\Big]
{1\over 4\pi}
\Big( R_1(|{\vec k}|){\Big)}^2
\ {(k^0+m)^2\over (k^++m)^2+{\vec k}_{\perp}^2}\
{1\over |{\vec k}|^2}\Big[ (k^+-m)^2-{\vec k}_{\perp}^2\Big] \ ,
\nonumber\\
h_1(x)&=&
\int \Big[ {\mathrm d}^2{\vec k}_{\perp}\Big]
{1\over 4\pi}
\Big( R_1(|{\vec k}|){\Big)}^2
\ {(k^0+m)^2\over (k^++m)^2+{\vec k}_{\perp}^2}\
{1\over |{\vec k}|^2}\Big[ - (k^+-m)^2\Big] \ ,
\nonumber\\
g^{\rm BD}_1(x)&=&
\int \Big[ {\mathrm d}^2{\vec k}_{\perp}\Big]
{1\over 4\pi}
\Big( R_1(|{\vec k}|){\Big)}^2
%\ {1\over |{\vec k}|^2}\Big[ (k^3)^2-{\vec k}_{\perp}^2\Big]
\ \Big[ - \, {1\over 3}\ \Big]
\ .
\label{k4}
\end{eqnarray}
We checked by explicit calculation
that $h^{\rm BD}_1(x,{\vec k}_{\perp})$ is the same as
$g^{\rm BD}_1(x,{\vec k}_{\perp})$ given in (\ref{k4}).
{}From the results in (\ref{k4}), we see that the Soffer's
inequality is satisfied with equality in this model:
$f_1(x)+g_1(x)=2|h_1(x)|$.

\section{Conclusion}

When we define $g_1^{\rm BD}(x)$ as the probability of the quark's being
in the $u_1^{\rm BD}(k)$ state minus that of being in the
$u_2^{\rm BD}(k)$ state when the proton's state is $|\lambda =+{1\over 2}>$,
and define $h_1^{\rm BD}(x)$ as the probability of the quark's being
in the ${1\over {\sqrt{2}}}(u_1^{\rm BD}(k)+u_2^{\rm BD}(k))$ state minus
that of being in the ${1\over {\sqrt{2}}}(u_1^{\rm BD}(k)-u_2^{\rm BD}(k))$ state
when the proton's state is ${1\over {\sqrt{2}}}(|\lambda =+{1\over 2}>+|\lambda =-{1\over 2}>)$,
the relation $g_1^{\rm BD}(x)=h_1^{\rm BD}(x)$ is satisfied.
The reason for the above is the following: $u_1^{\rm BD}(k)$, $u_2^{\rm BD}(k)$ and
$|\lambda =+{1\over 2}>$ are all angular momentum eigenstates of spin-half with the
eigenvalues of $j^3$ (or $J^3$) as $+{1\over 2}$ or $-{1\over 2}$,
and ${1\over {\sqrt{2}}}(u_1^{\rm BD}(k)+u_2^{\rm BD}(k))$,
${1\over {\sqrt{2}}}(u_1^{\rm BD}(k)-u_2^{\rm BD}(k))$ and
${1\over {\sqrt{2}}}(|\lambda =+{1\over 2}>+|\lambda =-{1\over 2}>)$ are
all angular momentum eigenstates of spin-half with the eigenvalues of $j^2$ (or $J^2$) as $+{1\over 2}$
or $-{1\over 2}$.
%and the latter three states can be obtained by rotating the former three states by rotating 90 degrees
%along the positive $y$ axis.
Therefore, the former and latter three states are equivalent and
only their quantization axes are different.
Then, the relation $g_1^{\rm BD}(x)=h_1^{\rm BD}(x)$ is satisfied.

However, the situation concerning the relation between the helicity and transversity
distributions $g_1(x)$ and $h_1(x)$ is different.
The states given by ${1\over {\sqrt{2}}}(u_1^{\rm LF}(k)\pm u_2^{\rm LF}(k))$ are not
angular momentum eigenstates of spin-half with the eigenvalue of $j^2$ as $\pm {1\over 2}$,
and there is no equivalence which existed in the previous paragraph for
$g_1^{\rm BD}(x)$ and $h_1^{\rm BD}(x)$.
Then, $g_1(x)$ and $h_1(x)$ are not equal.
The condition of $g_1(x)$ and $h_1(x)$ being equal is that the quark
transverse momentum is zero.
We explained these properties and also showed that $g_1(x)$ is a helicity distribution
only when the quark mass is zero.

\section*{Acknowledgements}
This work was supported in part by the International Cooperation
Program of the KICOS (Korea Foundation for International Cooperation
of Science \& Technology).

\vfill\pagebreak

\section*{Appendix A}
%: $u^{BD}(p)$, $u^{LC}(p)$ and
%$u^{{\vec{p}}\cdot {\vec{\Sigma}}\over |{\vec{p}}|}(p)$}

\section*{A1 $\ $ Relation between $u^{BD}(p)$ and $u^{LF}(p)$}

We use the $\gamma$ matrices in the Dirac representation:
\begin{eqnarray}
&&\gamma^0=\left(
\begin{array}{cc}
1&0\\
0&-1
\end{array}
\right)
\ ,\
\gamma^i=\left(
\begin{array}{cc}
0&{\sigma}^i\\
-{\sigma}^i&0
\end{array}
\right)
\ ,\
\nonumber\\
&&\gamma_5=\left(
\begin{array}{cc}
0&1\\
1&0
\end{array}
\right)
\ ,\
\alpha^i=\gamma^0\gamma^i=\left(
\begin{array}{cc}
0&{\sigma}^i\\
{\sigma}^i&0
\end{array}
\right)
\ ,\
\sigma^{12}=
\left(
\begin{array}{cc}
{\sigma}^3&0\\
0&{\sigma}^3
\end{array}
\right)
\ ,\
\nonumber\\
&&
\sigma^1=\left(
\begin{array}{cc}
0&1\\
1&0
\end{array}
\right)
\ ,\
\sigma^2=\left(
\begin{array}{cc}
0&-i\\
i&0
\end{array}
\right)
\ ,\
\sigma^3=\left(
\begin{array}{cc}
1&0\\
0&-1
\end{array}
\right)\ .
\label{a1app}
\end{eqnarray}

We use the following notations:
\begin{equation}
p^R=p^1+ip^2\ ,\ \
p^L=p^1-ip^2\ ,\qquad
p^+=p^0+p^3\ ,\ \
p^-=p^0-p^3\ .
\label{s3app}
\end{equation}

Let us study the equation
\begin{equation}
({\not{p}}-m)\ u(p)\ =0\ .
\label{a2app}
\end{equation}

The following $u(p)$ satisfies (\ref{a2app}):
\begin{equation}
u(p)\ =\ {1\over {\sqrt{N}}}\ ({\not{p}}+m)\ \gamma^0\ \chi\ .
\label{a3app}
\end{equation}

\subsection*{A1.1 $\ $ $u^{BD}(p)$}

%\subsection{$u^{BD}(p)$}

When we take $\chi$ and $N$ as
\begin{equation}
\chi^{BD}_1=\left(
\begin{array}{c}
1\\
0\\
0\\
0
\end{array}
\right)
\ ,\qquad
\chi^{BD}_2=\left(
\begin{array}{c}
0\\
1\\
0\\
0
\end{array}
\right)
\ ,\qquad
N^{BD}=p^0+m\ ,
\label{a4app}
\end{equation}
we have two linearly independent solutions:
\begin{equation}
u_i^{BD}(p)={{\not{p}}+m\over {\sqrt{p^0+m}}}\chi^{BD}_i
=\left(
\begin{array}{c}
{\sqrt{p^0+m}}\ \phi^{BD}_i\\
{{\vec{p}}\cdot {\vec{\sigma}}\over {\sqrt{p^0+m}}}\ \phi^{BD}_i
\end{array}
\right)\ ,
\label{a6app}
\end{equation}
where
\begin{equation}
\phi^{BD}_1=\left(
\begin{array}{c}
1\\
0
\end{array}
\right)\ ,\qquad
\phi^{BD}_2=\left(
\begin{array}{c}
0\\
1
\end{array}
\right)\ .
\label{a7app}
\end{equation}
When we write (\ref{a6app}) explicitly, we have
\begin{equation}
u_1^{BD}(p)={1\over {\sqrt{p^0+m}}}
\left(
\begin{array}{c}
p^0+m\\
0\\
p^3\\
p^R
\end{array}
\right)
\ ,
\qquad
u_2^{BD}(p)={1\over {\sqrt{p^0+m}}}
\left(
\begin{array}{c}
0\\
p^0+m\\
p^L\\
-p^3
\end{array}
\right)
\ .
\label{s1app}
\end{equation}

\subsection*{A1.2 $\ $  $u^{LF}(p)$}

%\subsection{$u^{LF}(p)$}

When we take $\chi$ and $N$ as
\begin{equation}
\chi^{LF}_1={1\over {\sqrt{2}}}\left(
\begin{array}{c}
1\\
0\\
1\\
0
\end{array}
\right)
\ ,\qquad
\chi^{LF}_2={1\over {\sqrt{2}}}\left(
\begin{array}{c}
0\\
1\\
0\\
-1
\end{array}
\right)
\ ,\qquad
N^{LF}=p^+\ ,
\label{a4aapp}
\end{equation}
we have another set of two linearly independent solutions
of (\ref{a2app}):
\begin{equation}
u_i^{LF}(p)\ =\
{1\over {\sqrt{p^+}}}\
(p^++\beta m+{\vec{\alpha}}_\perp \cdot {\vec{p}}_\perp)\
\chi^{LF}_i\ .
\label{a5app}
\end{equation}
We adopt the convention of Ref. \cite{LB80} in (\ref{a5app}).
%(We adopt the convention of BL PRD 22 (1980) 2157.)
When we write (\ref{a5app}) explicitly, we have
\begin{equation}
u_1^{LF}(p)={1\over {\sqrt{2p^+}}}
\left(
\begin{array}{c}
p^++m\\
p^R\\
p^+-m\\
p^R
\end{array}
\right)
\ ,
\qquad
u_2^{LF}(p)={1\over {\sqrt{2p^+}}}
\left(
\begin{array}{c}
-p^L\\
p^++m\\
p^L\\
-p^++m
\end{array}
\right)
\ .
\label{s2app}
\end{equation}

The two different sets $u^{BD}(p)$ in (\ref{s1app}) and
$u^{LF}(p)$ in (\ref{s2app}) are related as:
\begin{eqnarray}
\left(
\begin{array}{c}
u_1^{BD}(p)\\
u_2^{BD}(p)
\end{array}
\right)&=&
{1\over {\sqrt{2p^+(p^0+m)}}}\
\left(
\begin{array}{cc}
(p^++m)&-p^R\\
p^L&(p^++m)
\end{array}
\right)\ 
\left(
\begin{array}{c}
u_1^{LF}(p)\\
u_2^{LF}(p)
\end{array}
\right)\ ,
\label{s6app}\\
\left(
\begin{array}{c}
u_1^{LF}(p)\\
u_2^{LF}(p)
\end{array}
\right)&=&
{1\over {\sqrt{2p^+(p^0+m)}}}\
\left(
\begin{array}{cc}
(p^++m)&p^R\\
-p^L&(p^++m)
\end{array}
\right)\
\left(
\begin{array}{c}
u_1^{BD}(p)\\
u_2^{BD}(p)
\end{array}
\right)\ .
\label{s7app}
\end{eqnarray}

\section*{A2 $\ $  $u^{{\vec{p}}\cdot {\vec{\Sigma}}\over |{\vec{p}}|}(p)$}

%\section{$u^{{\vec{p}}\cdot {\vec{\Sigma}}\over |{\vec{p}}|}(p)$}
The spin matrix is given by
\begin{equation}
\Sigma^i=
\left(
\begin{array}{cc}
\sigma^i&0\\
0&\sigma^i
\end{array}
\right)\ .
\label{s8app}
\end{equation}
When we write ${\vec{p}}\cdot {\vec{\Sigma}}$ and ${\not{p}}$ matrices explicitly,
we have
\begin{equation}
{\vec{p}}\cdot {\vec{\Sigma}}=
\left(
\begin{array}{cccc}
p^3&p^L&0&0\\
p^R&-p^3&0&0\\
0&0&p^3&p^L\\
0&0&p^R&-p^3
\end{array}
\right)\ ,
{\not{p}}=
\left(
\begin{array}{cccc}
p^0&0&-p^3&-p^L\\
0&p^0&-p^R&p^3\\
p^3&p^L&-p^0&0\\
p^R&-p^3&0&-p^0
\end{array}
\right)\ .
\label{s9app}
\end{equation}
We can check by explicit matrix multiplications of ${\vec{p}}\cdot {\vec{\Sigma}}$
and ${\not{p}}$ matrices in (\ref{s9app}) that
\begin{equation}
{\not{p}}\ ({\vec{p}}\cdot {\vec{\Sigma}})\ =\
({\vec{p}}\cdot {\vec{\Sigma}})\ {\not{p}}\ .
\label{s11app}
\end{equation}
Let us find the eigenstates of ${\vec{p}}\cdot {\vec{\Sigma}}$ which
satisfy
\begin{equation}
{\vec{p}}\cdot {\vec{\Sigma}}\ u(p)\ =\
\lambda\ u(p)\ .
\label{s12app}
\end{equation}
From $|{\vec{p}}\cdot {\vec{\Sigma}}\ -\ \lambda I|\ =\ 0$ we have
$\lambda =+|{\vec{p}}|$ and $\lambda =-|{\vec{p}}|$.

For $\lambda =+|{\vec{p}}|$, the solution of (\ref{s12app}) is given by
\begin{equation}
u_{+1}^{{\hat {\vec{p}}}\cdot {\vec{\Sigma}}}(p)\ =\
{1\over {\sqrt{2|{\vec{p}}|(|{\vec{p}}|+p^3)}}}\ 
\Bigl( \ (|{\vec{p}}|+p^3)\ u^{BD}_1(p)\ +\ p^R\ u^{BD}_2(p)\ \Bigr)\ ,
\label{s13app}
\end{equation}
which is given explicitly as
\begin{equation}
u_{+1}^{{\hat {\vec{p}}}\cdot {\vec{\Sigma}}}(p)
=
{1\over {\sqrt{2|{\vec{p}}|(|{\vec{p}}|+p^3)}}}\
{1\over {\sqrt{p^0+m}}}\
\left(
\begin{array}{c}
(p^0+m)
\left(
\begin{array}{c}
|{\vec{p}}|+p^3\\
p^R
\end{array}
\right)\\
|{\vec{p}}|
\left(
\begin{array}{c}
|{\vec{p}}|+p^3\\
p^R
\end{array}
\right)
\end{array}
\right)\ .
\label{s14app}
\end{equation}
Using (\ref{s6app}), (\ref{s13app}) can also be written as
\begin{eqnarray}
&&u_{+1}^{{\hat {\vec{p}}}\cdot {\vec{\Sigma}}}(p)
\nonumber\\
&=&
{1\over {\sqrt{2p^+(p^0+m)}}}\ {1\over {\sqrt{2|{\vec{p}}|(|{\vec{p}}|+p^3)}}}
\label{s13aapp}\\
&\times& \Bigl( \ (|{\vec{p}}|+p^3)\Bigl((p^++m)\ +\ (|{\vec{p}}|-p^3)\Bigr)
\ u^{LF}_1(p)\ +\
p^R\ \Bigl( (p^++m)\ -\ (|{\vec{p}}|+p^3)\Bigr)
\ u^{LF}_2(p)\ \Bigr)\ .
\nonumber
\end{eqnarray}
For reference, if we consider the case of $m=0$, (\ref{s13aapp}) becomes
\begin{equation}
u_{+1}^{{\hat {\vec{p}}}\cdot {\vec{\Sigma}}}(p\ ;\ m=0)
\ =\ u^{LF}_1(p)\ .
\label{s13bapp}
\end{equation}
We chose the phase of
$u_{+1}^{{\hat {\vec{p}}}\cdot {\vec{\Sigma}}}(p)$ so that (\ref{s13bapp}) is
satisfied with identity.

For $\lambda =-|{\vec{p}}|$, the solution of (\ref{s12app}) is given by
\begin{equation}
u_{-1}^{{\hat {\vec{p}}}\cdot {\vec{\Sigma}}}(p)\ =\
{1\over {\sqrt{2|{\vec{p}}|(|{\vec{p}}|+p^3)}}}\
\Bigl(\ - \ p^L\ u^{BD}_1(p)\ +\ ( |{\vec{p}}|+p^3)\ u^{BD}_2(p)\ \Bigr)\ ,
\label{s15app}
\end{equation}
which is given explicitly as
\begin{equation}
u_{-1}^{{\hat {\vec{p}}}\cdot {\vec{\Sigma}}}(p)
=
{1\over {\sqrt{2|{\vec{p}}|(|{\vec{p}}|+p^3)}}}\
{1\over {\sqrt{p^0+m}}}\
\left(
\begin{array}{c}
(p^0+m)
\left(
\begin{array}{c}
-p^L\\
|{\vec{p}}|+p^3
\end{array}
\right)\\
-|{\vec{p}}|
\left(
\begin{array}{c}
-p^L\\
|{\vec{p}}|+p^3
\end{array}
\right)
\end{array}
\right)\ .
\label{s16app}
\end{equation}
Using (\ref{s6app}), (\ref{s15app}) can also be written as
\begin{eqnarray}
&&u_{-1}^{{\hat {\vec{p}}}\cdot {\vec{\Sigma}}}(p)
\nonumber\\
&=&
{1\over {\sqrt{2p^+(p^0+m)}}}\ {1\over {\sqrt{2|{\vec{p}}|(|{\vec{p}}|+p^3)}}}
\label{s15aapp}\\
&\times& \Bigl( \ -\ p^L\ \Bigl( (p^++m)\ -\ (|{\vec{p}}|+p^3)\Bigr)
\ u^{LF}_1(p)\ +\
(|{\vec{p}}|+p^3)\Bigl((p^++m)\ +\ (|{\vec{p}}|-p^3)\Bigr)
\ u^{LF}_2(p)\ \Bigr)\ .
\nonumber
\end{eqnarray}
For reference, if we consider the case of $m=0$, (\ref{s15aapp}) becomes
\begin{equation}
u_{-1}^{{\hat {\vec{p}}}\cdot {\vec{\Sigma}}}(p\ ;\ m=0)
\ =\ 
u^{LF}_2(p)\ .
\label{s15bapp}
\end{equation}

From (\ref{s13app}) and (\ref{s15app}) we get
\begin{eqnarray}
u_1^{BD}(p)&=&
{1\over {\sqrt{2|{\vec{p}}|(|{\vec{p}}|+p^3)}}}\
\Bigl( \ (|{\vec{p}}|+p^3)
\ u_{+1}^{{\hat {\vec{p}}}\cdot {\vec{\Sigma}}}(p)
\ -\
p^R
\ u_{-1}^{{\hat {\vec{p}}}\cdot {\vec{\Sigma}}}(p)
\ \Bigr)\ ,
\label{s17app}\\
u_2^{BD}(p)&=&
{1\over {\sqrt{2|{\vec{p}}|(|{\vec{p}}|+p^3)}}}\
\Bigl( \ p^L\ u_{+1}^{{\hat {\vec{p}}}\cdot {\vec{\Sigma}}}(p)
\ +\
(|{\vec{p}}|+p^3)
\ u_{-1}^{{\hat {\vec{p}}}\cdot {\vec{\Sigma}}}(p)
\ \Bigr)\ ,
\nonumber
\end{eqnarray}
and from (\ref{s13aapp}) and (\ref{s15aapp}) we get
\begin{eqnarray}
&&u_1^{LF}(p)
\label{s18app}\\
&=&
{1\over {\sqrt{2p^+(p^0+m)}}}\ {1\over {\sqrt{2|{\vec{p}}|(|{\vec{p}}|+p^3)}}}
\nonumber\\
&\times& \Bigl( \ 
(|{\vec{p}}|+p^3)\Bigl( (p^++m)\ +\ (|{\vec{p}}|-p^3)\Bigr)
\ u_{+1}^{{\hat {\vec{p}}}\cdot {\vec{\Sigma}}}(p)\ -\
p^R\ \Bigl( (p^++m)\ -\ (|{\vec{p}}|+p^3)\Bigr)
\ u_{-1}^{{\hat {\vec{p}}}\cdot {\vec{\Sigma}}}(p) \ \Bigr)\ ,
\nonumber\\
&&u_2^{LF}(p)
\nonumber\\
&=&
{1\over {\sqrt{2p^+(p^0+m)}}}\ {1\over {\sqrt{2|{\vec{p}}|(|{\vec{p}}|+p^3)}}}
\nonumber\\
&\times& \Bigl( \
p^L\ \Bigl( (p^++m)\ -\ (|{\vec{p}}|+p^3)\Bigr)
\ u_{+1}^{{\hat {\vec{p}}}\cdot {\vec{\Sigma}}}(p)\ +\
(|{\vec{p}}|+p^3)\Bigl( (p^++m)\ +\ (|{\vec{p}}|-p^3)\Bigr)
\ u_{-1}^{{\hat {\vec{p}}}\cdot {\vec{\Sigma}}}(p) \ \Bigr)\ .
\nonumber
\end{eqnarray}
%\\

%\vfill

%\pagebreak

\section*{A3 $\ $  Unitary Matrices}

We can write the relations among $u^{BD}(p)$, $u^{LF}(p)$ and
$u^{{\vec{p}}\cdot {\vec{\Sigma}}\over |{\vec{p}}|}(p)$ by unitary matrices as follows:
\begin{equation}
\left(
\begin{array}{c}
u_1^{BD}(p)\\
u_2^{BD}(p)
\end{array}
\right)\ =\
{\cal U}^{-1}\ 
\left(
\begin{array}{c}
u_1^{LF}(p)\\
u_2^{LF}(p)
\end{array}
\right) \ ,
\label{s7app}
\end{equation}
where
\begin{equation}
{\cal U}^{-1}\ =\
{1\over {\sqrt{2p^+(p^0+m)}}}\
\left(
\begin{array}{cc}
p^++m&-p^R\\
p^L&p^++m
\end{array}
\right)
\ ,
\label{s7mapp}
\end{equation}
\begin{equation}
{\cal U}\ =\
{1\over {\sqrt{2p^+(p^0+m)}}}\
\left(
\begin{array}{cc}
p^++m&p^R\\
-p^L&p^++m
\end{array}
\right)\ ,
\label{s7mmapp}
\end{equation}
\begin{equation}
\left(
\begin{array}{c}
u_{+1}^{{\hat {\vec{p}}}\cdot {\vec{\Sigma}}}(p)\\
u_{-1}^{{\hat {\vec{p}}}\cdot {\vec{\Sigma}}}(p)
\end{array}
\right)\ =\
{\cal V}\
\left(
\begin{array}{c}
u_1^{BD}(p)\\
u_2^{BD}(p)
\end{array}
\right) \ ,
\label{m3app}
\end{equation}
where
\begin{equation}
{\cal V}\ =\
{1\over {\sqrt{2|{\vec{p}}|(|{\vec{p}}|+p^3)}}}\
\left(
\begin{array}{cc}
|{\vec{p}}|+p^3&p^R\\
-p^L&|{\vec{p}}|+p^3
\end{array}
\right)\ ,
\label{m4app}
\end{equation}
and
\begin{equation}
\left(
\begin{array}{c}
u_{+1}^{{\hat {\vec{p}}}\cdot {\vec{\Sigma}}}(p)\\
u_{-1}^{{\hat {\vec{p}}}\cdot {\vec{\Sigma}}}(p)
\end{array}
\right)\ =\
{\cal W}\
\left(
\begin{array}{c}
u_1^{LF}(p)\\
u_2^{LF}(p)
\end{array}
\right) \ ,
\label{m5app}
\end{equation}
where
\begin{eqnarray}
&&{\cal W}\ =\
{1\over {\sqrt{2p^+(p^0+m)}}}\ {1\over {\sqrt{2|{\vec{p}}|(|{\vec{p}}|+p^3)}}}
\label{m6app}\\
&\times&
\left(
\begin{array}{cc}
(|{\vec{p}}|+p^3)\Bigl((p^++m)\ +\ (|{\vec{p}}|-p^3)\Bigr)&
p^R\ \Bigl( (p^++m)\ -\ (|{\vec{p}}|+p^3)\Bigr)\\
-p^L\ \Bigl( (p^++m)\ -\ (|{\vec{p}}|+p^3)\Bigr)&
(|{\vec{p}}|+p^3)\Bigl((p^++m)\ +\ (|{\vec{p}}|-p^3)\Bigr)
\end{array}
\right)\ .
\nonumber
\end{eqnarray}
We can check that the relation ${\cal W}\ =\ {\cal V}\ {\cal U}^{-1}$
is satisfied.
%\begin{equation}
%{\cal W}\ =\ {\cal V}\ {\cal U}^{-1}\ .
%\label{m7app}
%\end{equation}
%\\

We can express the relations in the above as follows:
\begin{equation}
\left(
\begin{array}{c}
u_1^{BD}\\
u_2^{BD}
\end{array}
\right)\ =\
{\cal U}^{-1}\
\left(
\begin{array}{c}
u_1^{LF}\\
u_2^{LF}
\end{array}
\right)
\ ,\ \
\left(
\begin{array}{c}
u_{+1}^{{\hat {\vec{p}}}\cdot {\vec{\Sigma}}}\\
u_{-1}^{{\hat {\vec{p}}}\cdot {\vec{\Sigma}}}
\end{array}
\right)\ =\
{\cal V}\
\left(
\begin{array}{c}
u_1^{BD}\\
u_2^{BD}
\end{array}
\right)
\ ,\ \
\left(
\begin{array}{c}
u_{+1}^{{\hat {\vec{p}}}\cdot {\vec{\Sigma}}}\\
u_{-1}^{{\hat {\vec{p}}}\cdot {\vec{\Sigma}}}
\end{array}
\right)\ =\
{\cal W}\
\left(
\begin{array}{c}
u_1^{LF}\\
u_2^{LF}
\end{array}
\right)
\ ,
\label{mmm1app}
\end{equation}
where
\begin{equation}
{\cal U}\ =\
{1\over {\sqrt{2p^+(p^0+m)}}}\
\left(
\begin{array}{cc}
p^++m&p^R\\
-p^L&p^++m
\end{array}
\right)\ =\
\left(
\begin{array}{cc}
{\rm cos}\theta&{\rm sin}\theta\ e^{i\phi}\\
-\ {\rm sin}\theta\ e^{-i\phi}&{\rm cos}\theta
\end{array}
\right) \ ,
\label{m8app}
\end{equation}
\begin{equation}
{\rm cos}\theta\ =\
{p^++m\over {\sqrt{2p^+(p^0+m)}}}\ ,
\qquad
{\rm sin}\theta\ =\
{|{\vec p}_\perp |\over {\sqrt{2p^+(p^0+m)}}}\ ,
\qquad
|{\vec p}_\perp | \ =\ {\sqrt{(p^1)^2+(p^2)^2}}\ ,
\label{m9app}
\end{equation}
\begin{equation}
e^{i\phi}\ =\ {p^R\over |{\vec p}_\perp |}\ ,\qquad
e^{-i\phi}\ =\ {p^L\over |{\vec p}_\perp |}\ ,
\label{m10app}
\end{equation}
\begin{equation}
{\cal V}\ =\
{1\over {\sqrt{2|{\vec{p}}|(|{\vec{p}}|+p^3)}}}\
\left(
\begin{array}{cc}
|{\vec{p}}|+p^3&p^R\\
-p^L&|{\vec{p}}|+p^3
\end{array}
\right)\ =\
\left(
\begin{array}{cc}
{\rm cos}\chi&{\rm sin}\chi\ e^{i\phi}\\
-\ {\rm sin}\chi\ e^{-i\phi}&{\rm cos}\chi
\end{array}
\right) \ ,
\label{m11app}
\end{equation}
\begin{equation}
{\rm cos}\chi\ =\
{|{\vec{p}}|+p^3\over {\sqrt{2|{\vec{p}}|(|{\vec{p}}|+p^3)}}}\ ,
\qquad
{\rm sin}\chi\ =\
{|{\vec p}_\perp |\over {\sqrt{2|{\vec{p}}|(|{\vec{p}}|+p^3)}}}\ ,
\label{m12app}
\end{equation}
and
\begin{eqnarray}
&&{\cal W}\ =\
{1\over {\sqrt{2p^+(p^0+m)}}}\ {1\over {\sqrt{2|{\vec{p}}|(|{\vec{p}}|+p^3)}}}
\label{m13app}\\
&\times&
\left(
\begin{array}{cc}
(|{\vec{p}}|+p^3)\Bigl((p^++m)\ +\ (|{\vec{p}}|-p^3)\Bigr)&
p^R\ \Bigl( (p^++m)\ -\ (|{\vec{p}}|+p^3)\Bigr)\\
-p^L\ \Bigl( (p^++m)\ -\ (|{\vec{p}}|+p^3)\Bigr)&
(|{\vec{p}}|+p^3)\Bigl((p^++m)\ +\ (|{\vec{p}}|-p^3)\Bigr) \ ,
\end{array}
\right)
\nonumber\\
&=&
\left(
\begin{array}{cc}
{\rm cos}(\chi -\theta )&{\rm sin}(\chi -\theta )\ e^{i\phi}\\
-\ {\rm sin}(\chi -\theta )\ e^{-i\phi}&{\rm cos}(\chi -\theta )
\end{array}
\right)
\ =\
\left(
\begin{array}{cc}
{\rm cos}\psi&{\rm sin}\psi\ e^{i\phi}\\
-\ {\rm sin}\psi\ e^{-i\phi}&{\rm cos}\psi
\end{array}
\right)
\ 
\nonumber
\end{eqnarray}
\begin{eqnarray}
{\rm cos}\psi &=&
{(|{\vec{p}}|+p^3)\Bigl((p^++m)\ +\ (|{\vec{p}}|-p^3)\Bigr)\over
{\sqrt{2p^+(p^0+m)}}\ {\sqrt{2|{\vec{p}}|(|{\vec{p}}|+p^3)}}}\ ,
\label{m14app}\\
{\rm sin}\psi &=&
{|{\vec p}_\perp |\ \Bigl( (p^++m)\ -\ (|{\vec{p}}|+p^3)\Bigr)\over 
{\sqrt{2p^+(p^0+m)}}\ {\sqrt{2|{\vec{p}}|(|{\vec{p}}|+p^3)}}}\ .
\nonumber
\end{eqnarray}

We can write the relations in the above more compactly as:
\begin{equation}
{\cal U}\ =\
\left(
\begin{array}{cc}
{\rm cos}\theta&{\rm sin}\theta\ e^{i\phi}\\
-\ {\rm sin}\theta\ e^{-i\phi}&{\rm cos}\theta
\end{array}
\right)\ =\
I\ {\rm cos}\theta \ +\ i\ {\rm sin}\theta\ {\vec \sigma}\cdot {\hat n}
\ =\
e^{i\ {\vec \sigma}\cdot {\hat n}\ \theta} \ ,
\label{m15app}
\end{equation}
where
\begin{equation}
{\hat n}\ =\ (\ {\rm sin}\phi\ ,\ {\rm cos}\phi\ ,\ 0\ )\ ,
\label{m16app}
\end{equation}
\begin{eqnarray}
{\cal V}&=&
\left(
\begin{array}{cc}
{\rm cos}\chi&{\rm sin}\chi\ e^{i\phi}\\
-\ {\rm sin}\chi\ e^{-i\phi}&{\rm cos}\chi
\end{array}
\right)\ =\
e^{i\ {\vec \sigma}\cdot {\hat n}\ \chi} \ ,
\label{m17app}\\
{\cal W}&=&
\left(
\begin{array}{cc}
{\rm cos}\psi&{\rm sin}\psi\ e^{i\phi}\\
-\ {\rm sin}\psi\ e^{-i\phi}&{\rm cos}\psi
\end{array}
\right)\ =\
e^{i\ {\vec \sigma}\cdot {\hat n}\ \psi} \ .
\label{m18app}
\end{eqnarray}
The above expressions diven in (\ref{m15app}), (\ref{m17app}) and (\ref{m18app})
are useful.
For example, we can understand the relations written in the last line of (\ref{m13app})
easily as
\begin{equation}
{\cal W}={\cal V}\ {\cal U}^{-1}\ =\
e^{i\ {\vec \sigma}\cdot {\hat n}\ \chi}\
e^{i\ {\vec \sigma}\cdot {\hat n}\ (-\theta)}
\ =\ e^{i\ {\vec \sigma}\cdot {\hat n}\ (\chi -\theta)}\ .
\label{m19app}
\end{equation}

\section*{Appendix B}

The Bjorken-Drell spinors $u_i^{BD}(p)$ given in (\ref{a6app}) can be obtained
by applying the Lorentz boost operator $S(p)$ to the spinors $\chi_i^{BD}$
given in (\ref{a4app}), which is the positive enegy eigenstates of the Dirac equation
in the quark rest frame, where $S(p)$ is given by
\begin{equation}
S(p)=\sqrt{p^0+m}
\left(
\begin{array}{cc}
1&{{\vec p}\cdot {\vec \sigma}\over p^0+m}\\
{{\vec p}\cdot {\vec \sigma}\over p^0+m}&1
\end{array}
\right)
\ .
\label{appb1app}
\end{equation}
The the total angular momentum operator $j^i$ of quark is given by the sum of
the spin operator $s^i$ and the orbital angular momentum operator $l^i$ as:
\begin{equation}
j^i = s^i + l^i\ ,\ \ \
s^i={1\over 2}\, \Sigma^i =
{1\over 2}
\left(
\begin{array}{cc}
\sigma^i&0\\
0&\sigma^i
\end{array}
\right)
\ ,\ \ \
l^i = -i \epsilon^{ijk} k^j{\partial\over \partial k^k}\ ,\ \ \
\epsilon^{123}=1\ .
\label{p5appbapp}
\end{equation}

We can show that $j^i$ in (\ref{p5appbapp}) and $S(p)$ in (\ref{appb1app}) commute:
\begin{equation}
[j^i , S(p)]=0\ .
\label{appb3app}
\end{equation}
Then, if a quark spinor is an eigenstate of the operator $j^i$ in the quark
rest frame (in this frame $j^i=s^i$), the Bjorken-Drell spinor which is made by Lorentz
boosting that quark spinor by multiplying $S(p)$ given in (\ref{appb1app})
is also an eigenstate of the same operator $j^i$.
We illustrate this by considering an example in which the quark spinor is an eigenstate
of $s^2$ in the quark rest frame:
\begin{eqnarray}
&&j^2\, {1\over \sqrt{2}}\Big( u_1^{BD}(p)+u_2^{BD}(p)\Big)=
j^2\, S(p)\, {1\over \sqrt{2}}\Big( \chi_1^{BD}+\chi_2^{BD}\Big)=
S(p)\, j^2\, {1\over \sqrt{2}}\Big( \chi_1^{BD}+\chi_2^{BD}\Big)
\nonumber\\
&&
=S(p)\, s^2\, {1\over \sqrt{2}}\Big( \chi_1^{BD}+\chi_2^{BD}\Big)=
S(p)\, {1\over 2}\, {1\over \sqrt{2}}\Big( \chi_1^{BD}+\chi_2^{BD}\Big)=
{1\over 2}\, {1\over \sqrt{2}}\Big( u_1^{BD}(p)+u_2^{BD}(p)\Big) \ ,
\qquad
\label{appb4app}
\end{eqnarray}
where $\chi_i^{BD}$ and $u_i^{BD}(p)$ are given in (\ref{a4app}) and (\ref{a6app}).
Eq. (\ref{appb4app}) shows that ${1\over \sqrt{2}}\Big( \chi_1^{BD}+\chi_2^{BD}\Big)$
is the eigenstate of $s^2$ with eigenvalue $+{1\over 2}$ in the quark rest frame, and then
${1\over \sqrt{2}}\Big( u_1^{BD}(p)+u_2^{BD}(p)\Big)$ is the eigenstate of $j^2$
with eigenvalue $+{1\over 2}$.

\vfill\pagebreak

\end{document}